# Geometrical congruence and efficient greedy navigability of complex networks

Carlo Vittorio Cannistraci[1,2,*] and Alessandro Muscoloni[1]

[1] Biomedical Cybernetics Group, Biotechnology Center (BIOTEC), Center for Molecular and Cellular Bioengineering (CMCB), Center for Systems Biology Dresden (CSBD), Cluster of Excellence Physics of Life (PoL), Department of Physics, Technische Universität Dresden, Tatzberg 47/49, 01307 Dresden, Germany

[2] Center for Complex Network Intelligence (CCNI), Tsinghua Laboratory of Brain and Intelligence (THBI), Tsinghua University, 160 Chengfu Rd., SanCaiTang Building, Haidian District, 100084, Beijing, China

*Corresponding author: Carlo Vittorio Cannistraci (kalokagathos.agon@gmail.com)

**Hyperbolic networks are supposed to be congruent with their underlying latent geometry[1,2] and following geodesics in the hyperbolic space is believed equivalent to navigate through topological shortest paths (TSP)[1,2]. This assumption of geometrical congruence is considered the reason for nearly maximally efficient greedy navigation of hyperbolic networks[1,2].**

**Here, we propose a complex network measure termed *geometrical congruence* (GC) and we show that there might exist different TSP, whose projections (pTSP) in the hyperbolic space largely diverge, and significantly differ from the respective geodesics. We discover that, contrary to current belief, hyperbolic networks do not demonstrate in general geometrical congruence and efficient navigability which, in networks generated with nPSO model[3,4], seem to emerge only for power-law exponent close to 2. We conclude by showing that GC measure can impact also real networks analysis, indeed it significantly changes in structural brain connectomes grouped by gender or age.**

Studies in network geometry[2] suggest that the hidden geometrical space behind the observable topology of a network drives the navigation[5] of the network structure according to the distances between nodes in the latent geometrical space[6]. In particular, a recent pre-print of a review[2] that might soon become a crucial reference for the field of network geometry, reports a list of theoretical studies according to which *hyperbolic networks* are maximally efficient for such geometric navigation[1,7], and the main reason behind this phenomenon is assumed the proximity of TSP in the hyperbolic networks to the corresponding geodesics in the underlying hyperbolic geometry. If we enucleate better this concept, this means that the pTSP follow closely their associated hyperbolic geodesics in the latent space[1,2] and, since this is defined as geometrical congruence of the network topology with the hidden geometry, hyperbolic networks are believed geometrically congruent. Another key property for efficient geometrical navigation is the existence of superhubs that interconnect large parts of the network. This happens when the network degree distribution follows a power-law with exponent $\gamma<3$[1,2], in which case the networks are termed 'ultra-small-world'[2,8]. Boguñá et al.[9] proposed a theoretical demonstration that greedy navigation in networks with $\gamma<3$ and strong clustering (such as hyperbolic networks[1]) can always find these ultrashort paths which follow the geodesics[2,9], and thus navigation in hyperbolic networks with $\gamma<3$ is believed maximally efficient because of their supposed geometrical congruence[1,2]. Vice versa, a network that is maximally navigable by design is considered similar to hyperbolic networks[2].

Despite the abovementioned theoretical research, we have not found any study with computational evidences that validate these theoretical conclusions by means of numerical simulations that measure the level of congruence on hyperbolic network models. This might be due to the lack of definition of a computational measure of geometrical congruence in network science. Here, we address this problem and we proceed to measure geometrical congruence and greedy navigability in hyperbolic networks generated with the nonuniform popularity-similarity optimization (nPSO) model[3,4] which, in comparison to the classical PSO model[10], is a generalization that is able to grow realistic hyperbolic networks with tailored community structure.

Fig. 1 of this study starts with a 'bad news'. It represents a hyperbolic network generated with the nPSO model (parameters details in Fig. 1 legend). Fig. 1a displays the network in the hyperbolic disk where the links are drawn in grey colour according to the hyperbolic geodesics. Fig. 1b highlights in red colour the hyperbolic geodesic between two specific nonadjacent nodes in the network, in black colour the links involved in all the possible TSP between these two nodes and in green colour the geometrical shortest path (GSP, which in this specific case is unique). The GSP is the shortest path computed on the weighted network where the connections are weighted by the geometrical distances between the node pairs, therefore the length of the GSP is the sum of the geometrical distances of the connections involved in such path. Note that the pTSP is also a sum of the geometrical distances of the connections involved in the path, but in this case the path is the TSP, which is computed on the unweighted network. We clarify that if the GSP and the TSP have the same number of connections, then the length of the GSP is equivalent to the smallest pTSP. However, the GSP might also have a larger number of connections than the TSP and in that case there would not be equivalence. For instance, there might be a GSP as sum of 3 steps that in total are shorter than the projection of a TSP of 2 steps.

The bad news highlighted by the Fig. 1 is that for a certain geodesic we can have multiple topological shortest paths that visibly diverge from the geodesic path and this suggests that it would be important to appraise whether the distribution of geodesics between all pairwise nonadjacent nodes significantly differ from the distribution of the $\overline{pTSP}$ between the same nodes. Note that $\overline{pTSP}$ is the mean of the pTSP between two nodes, we take the mean as central measure because for each geodesic there might be multiple pTSP. Fig. 2 displays a comparison of these two distributions when we fix a parameter of the nPSO model and we vary the others. This helps to discuss how the congruence of geodesics and pTSP varies according to a certain specific structural property which is adopted to shape the hyperbolic network according to the nPSO model. Three structural properties are discussed: if average degree $\bar{d}$ grows (m is the direct nPSO model parameter, and $\bar{d}$ is about 2*m for sparse networks; we consider m = [2, 6, 10] hence $\bar{d}$ about [4, 12, 20]), network density increases; if temperature T grows (we consider T = [0.1, 0.5, 0.9]), clustering decreases; if γ grows (we consider γ = [2, 2.5, 3]), superhub structure of the network is mitigated. We keep N=100 because this network node size is enough to properly discuss these structural properties. For each panel of the figure the solid line indicates that community organization is not imposed (C=0, this is equivalent to the classical uniform PSO model[10] and the dashed line indicates that networks with 4 communities are considered (C=4). In general, these two different community organizations seem not relevant for our investigation and therefore they are not further discussed. Fig. 2a has three panels where from left to right the average degree increases while the other parameters are fixed to their intermediate value. It emerges that when average degree grows, $\overline{pTSP}$ between nodes shrinks and approximates better the geodesic whose distribution is unimodal. The distribution of $\overline{pTSP}$ is multimodal because each peak is associated to a different possible topological shortest path length. For small average degree equal to 4 the network is closer to a tree and the peaks of the distribution are associated with topological shortest paths of length 2, 3 and 4 respectively, with path 3 being the prevalent. When average degree is 12, the $\overline{pTSP}$ distribution has two peaks that are associated to TSP of length 2 and 3. When average degree is 20, there is a prevalence of TSP of length 2 because the network is quite dense. Fig. 2b has three panels where from left to right the power-law exponent γ increases while the other parameters are fixed to their intermediate value. It emerges that when γ grows, $\overline{pTSP}$ distribution remarkably diverges from the geodesic. For γ=2, $\overline{pTSP}$ distribution has a predominance of TSP of length 2 that are also very close to the geodesic. Whereas for γ=2.5 and γ=3 the trend is similar and consists of a bimodal distribution with a first peak for TSP of length 2 and a second for TSP of length 3, but in general these paths are less congruent with the geodesic. Fig. 2c has three panels where from left to right the temperature T increases while the other parameters are fixed to their intermediate value. It emerges that when T grows (it means that clustering decreases), $\overline{pTSP}$ distribution seems to maintain the same level of divergence from the geodesic distribution. This might mislead to the conclusion that clustering does not seem to impact the geometrical congruence between $\overline{pTSP}$ and geodesic. However, a closer investigation of Fig. 2c suggests that when T grows, $\overline{pTSP}$ distribution kurtosis is modified. This implies that in nPSO hyperbolic networks with higher clustering the $\overline{pTSP}$ values related to TSP of the same length are more congruent between them. Finally, we perform for each of the 9 subplots of Fig. 2 a Mann-Whitney statistical test to assess in which of these scenarios the geodesic distribution and $\overline{pTSP}$ distribution do not differ and we can accept the hypothesis of geometrical congruence between geodesics and associated $\overline{pTSP}$. Astonishingly, the result of the statistical test (considering p-value<0.05 as significance level) is that the hypothesis of congruence should be always rejected, geodesic and $\overline{pTSP}$ significantly differ for all possible investigated parameter combinations of the nPSO model hyperbolic networks. Therefore, the first important finding of this study is that we cannot statistically accept that hyperbolic networks in general are congruent with their latent geometry, which is the current credence in the scientific literature.

At this point of our study, we have to raise the level of the scientific precision adopted to investigate the hypothesis of geometrical congruence of hyperbolic networks, and to do so we have to invent novel network science tools that allow to dive deeper in the conceptual and methodical definition of geometrical congruence. We let you notice that current network science literature discusses of proximity and congruence of the geodesic to the pTSP in a qualitative and visual-based fashion that might leave space to misinterpretation and misunderstanding. Hence, a key innovation of this study is that we introduce a general measure of *geometrical congruence* in complex networks that will be fundamental to quantitatively evaluate the extent to which geometrical networks are congruent with their latent geometry. Given a network with *n* nodes and *e* edges, we define the geometrical congruence (GC) as:

$$GC(\overline{pTSP}, RD) = \left(\frac{2}{n \cdot (n-1) - 2 \cdot e}\right) \cdot \sum_{i<j} \frac{RD(i,j)}{\overline{pTSP}(i,j)} \; ; \; \text{with } (i,j) \in \tilde{E}$$

where $\tilde{E}$ is the set of pairs $(i,j)$ of nonadjacent nodes.

The computation of the $\overline{pTSP}(i,j)$ is nontrivial because it requires to find all the possible TSP between (i,j). The technical details on how to solve this computational problem are provided in the Methods section. RD(i,j) can be any node pairwise reference distance (not necessarily restricted to the geodesic). For instance, in this study we consider RD(i,j) equal to the geodesic (GEO) in one case and to the geometrical shortest path (GSP) in the second case. Hence, in the first case we measure the GC with the geodesic, in the second case we measure the GC with the GSP. Since a GC=0.5 means that on average in the network $\overline{pTSP}(i,j)$ is twice the length of RD(i,j) (indicating a low congruence) we consider the following definition of the scale of values for GC:

GC = [0, 0.4[ indicates negligible congruence; GC = [0.4, 0.6[ indicates low congruence; GC = [0.6, 0.8[ indicates medium congruence; GC = [0.8, 1] indicates high congruence.

Fig. 3 reports the values of GC measure and greedy navigability measure in hyperbolic networks generated with the nPSO model across different parameter combinations. In particular, Fig. 3a shows a heatmap with average GC($\overline{pTSP}$, GEO) on 10 realizations of nPSO hyperbolic networks (N=100, T=0.1, 4 communities) spanned across a large combination of average degree $\bar{d}$ and power-law exponent γ. From Fig. 3a emerges that high GC($\overline{pTSP}$, GEO) is reached in these hyperbolic networks only for γ=2 and, most importantly, the measure of GC($\overline{pTSP}$, GEO) is strongly matched with a measure of navigability (Fig. 3b) termed greedy routing efficiency[11,12] (GRE, whose range of values is between 0 and 1, see Methods for details). We computed GRE(pGRP, GEO) by comparing the projection of the greedy routing paths (pGRP) with the respective geodesics between pairs of nonadjacent nodes. For T=0.3 (see Suppl. Fig. 1-4) the nPSO model networks seem to retain similar congruence and navigability, however for T=0.5 when the clustering vanishes (and consequently hyperbolic geometry vanishes[1,3,10]) also the congruence and navigability are significantly affected, and this result is in accordance with previous conclusions[1]. Fig. 3a-b results are confirmed also for networks with N=1000 and no fixed community organization (see Suppl. Fig. 1-4). Hence, the second important finding of this study is that in general for γ = ]2,3] the hyperbolic networks generated with the nPSO model present medium to low congruence and greedy navigability, and that high congruence/navigability emerges only for γ proximal to 2 (at least in these nPSO hyperbolic networks). These computational findings are important because they significantly correct and refine the results of previous theoretical studies[1,9], which are then included in a review study on network geometry[2], according to which the greedy navigation in hyperbolic networks with γ<3 can always find ultrashort paths which follow the geodesics, and thus navigation in these hyperbolic networks is maximally efficient. This is not true for all hyperbolic networks, and in Fig. 3a-b (and Suppl. Fig. 1-4) we offer computational evidence based on GC and GRE that this is not true for PSO[10] and nPSO[3] hyperbolic networks.

Interestingly, if we change the reference distance from geodesic to GSP (the green line in Fig. 1b) then the measures GC($\overline{pTSP}$, GSP) and GRE(pGRP, GSP) show high congruence and high navigability for any parameter combination (Fig. 3c-d) at T=0.1. On one side, this result suggests that using the GSP as reference distance for GC and GRE is improper and badly posed if the goal is to offer evidence of geometrical congruence of a network with its latent geometry, in which case the geodesic should be used as reference. On the other side, the proposed GC measure seems properly designed, since it is strongly associated to GRE on hyperbolic networks with high clustering (T=0.1) and, as expected according to theory[1], when temperature increases (T=0.3 and T=0.5, see Suppl. Fig. 1-4), clustering decreases, hyperbolic geometry tend to vanish and GC($\overline{pTSP}$, GSP) has a tendency to deviate from GRE(pGRP, GSP) which becomes more evident for C=4 and N=1000 (Suppl. Fig. 4). Therefore GC($\overline{pTSP}$, GSP) can still be considered an interesting marker to compare differences of geometrical congruence between structural and weighted connectivity across complex networks for which the latent geometry is unknown. For instance, Fig. 4a reports the results of the GC($\overline{pTSP}$, GSP) analysis on 614 macroscale structural MRI brain connectomes of healthy resting-state individuals (age range 7-85 years old) from Faskowitz et al.[13] divided for gender (230 male versus 384 females), and we find that GC($\overline{pTSP}$, GSP) is significantly higher (p-value<0.001, Mann-Whitney test) in females. Fig. 4b reports the analysis on a subsample (N=438) of the same dataset divided by age range in a young group (7-30 years old) and an elderly group (55-85 years old), and we find that GC($\overline{pTSP}$, GSP) is significantly higher (p-value<0.001, Mann-Whitney test) in the elderly group.

Altogether these findings could have practical impact on real applications for the design and engineering of communications networks, such as Internet, or for quantitative investigation of the topological-geometrical coupling, which is a novel and promising measurable network feature that might be associated to the organization and functionality of brain connectomes or other complex networks.

## Methods

### Geometrical congruence (GC)

Given a network with *n* nodes and *e* edges, we define the geometrical congruence (GC) as:

$$GC(\overline{pTSP}, RD) = \left(\frac{2}{n \cdot (n-1) - 2 \cdot e}\right) \cdot \sum_{i<j} \frac{RD(i,j)}{\overline{pTSP}(i,j)} \; ; \; \text{with } (i,j) \in \tilde{E}$$

where $\tilde{E}$ is the set of pairs (i,j) of nonadjacent nodes.

In practice, for each pair (i,j) of nonadjacent nodes we compute the ratio between RD(i,j), which is a reference distance, and $\overline{pTSP}(i,j)$, which is the mean projection of all the TSP between (i,j). The GC is obtained as the average of such ratios, assuming values between 0 and 1. Note that RD(i,j)=RD(j,i) and $\overline{pTSP}(i,j)=\overline{pTSP}(j,i)$, therefore we only evaluate for pairs (i,j) such that i<j. RD(i,j) can be any node pairwise reference distance. In this study, in one case we consider RD(i,j) equal to the geodesic (GEO) and in another case equal to the geometrical shortest path (GSP).

In the case of GEO, in our analysis they correspond to the pairwise hyperbolic distances between the nodes in the hyperbolic disk, which are provided in output by the nPSO model[3,4] when generating a network (see Methods section related to the nPSO model).

In the case of GSP, they correspond to the weighted shortest paths computed using the Johnson's algorithm[14] on the weighted network, where the connections are weighted by distances between the node pairs. If known, such distances can correspond to the geodesics, otherwise the weights can also represent other types of distances. The length of the GSP is equal to the sum of the distances of the connections involved in such path.

The computation of the $\overline{pTSP}(i,j)$ requires to find all the possible TSP between (i,j). First of all, we apply the Johnson's algorithm[14] on the unweighted network to obtain the length of the TSP for all the node pairs (this is a one-time computation, not needed for each pair individually). Then, for a given pair (i,j), we apply the Suppl. Algorithm 1 to find all the possible paths between (i,j) of length TSP(i,j). For each of these paths the projection (pTSP) is given by the sum of the weights (distances) of the connections involved in the path, and the $\overline{pTSP}(i,j)$ is finally obtained as the mean of the projections.

## Greedy routing efficiency (GRE)

Given a network with *n* nodes and *e* edges, the greedy routing efficiency (GRE)[11,12] in respect to the set $\tilde{E}$ of nonadjacent node pairs (i,j) is:

$$GRE(pGRP, RD) = \left(\frac{1}{n \cdot (n-1) - 2 \cdot e}\right) \cdot \sum \frac{RD(i,j)}{pGRP(i,j)} \ ; \ \text{with } (i,j) \in \tilde{E}$$

In practice, for each pair (i,j) of nonadjacent nodes we compute the ratio between RD(i,j), which is a reference distance, and pGRP(i,j), which is the projection of the greedy routing path between (i,j). The GRE is obtained as the average of such ratios, assuming values between 0 and 1. Note that pGRP(i,j) can be different from pGRP(j,i), therefore we evaluate both pairs (i,j) and (j,i). RD(i,j) can be any node pairwise reference distance. In this study, in one case we consider RD(i,j) equal to the geodesic (GEO) and in another case equal to the geometrical shortest path (GSP), for more details please refer to the previous section on geometrical congruence (GC).

In the measure of greedy routing efficiency (GRE) introduced in the original publications[11,12], all the node pairs have been considered, both adjacent and nonadjacent nodes, and the associated formula is:

$$GRE(pGRP, RD) = \left(\frac{1}{n \cdot (n-1)}\right) \cdot \sum \frac{RD(i,j)}{pGRP(i,j)} \ ; \ \text{with } (i,j) \in E$$

where $E$ is the set of pairs (i,j) of both adjacent and nonadjacent nodes. We note that in the GRE formulation adopted in this study only nonadjacent node pairs have been considered, in order to guarantee a fair comparison of the values with respect to the measure of geometrical congruence (GC), which is also evaluated only for nonadjacent node pairs. In addition, we clarify that in the previous publications[11,12], in which we originally introduce GRE, the formula was given only for the special case that pGRP is unweighted (hence it is the number of the greedy routing hops between two nodes) and RD=TSP. Here instead we propose for the first time a more general formula based on any reference distance that, according to the context of the scientific study, can be adequately selected.

In the algorithm to find the greedy routing path (GRP)[5] between a node pair (i,j), a packet is sent from the source i to the destination j. Every node knows only the geometrical coordinates of its neighbours and the ones of the destination j, which are written in the packet. At each step (hop) the packet is forwarded from the current node to its neighbour closest to the destination, meaning at the lowest geometrical distance (hyperbolic distance, in this study). The packet is dropped when this neighbour is the same from which the packet has been received at the previous step, since a loop has been generated, and the greedy routing is unsuccessful (pGRP is set to infinite). Otherwise, if the packet reaches the destination, the projection of the greedy routing path (pGRP) is equal to the sum of the distances of the connections involved in such path. In case of unweighted networks, this is equivalent to the number of hops that separate the node pair (i,j).

## Nonuniform popularity-similarity optimization model (nPSO) and classical popularity-similarity optimization model (PSO)

The Popularity-Similarity-Optimization (PSO) model[10] is a recently introduced generative model for networks that is based on growing soft random geometric graphs in the hyperbolic space. In this model the networks evolve optimizing a trade-off between node popularity (abstracted by the radial coordinate) and similarity (represented by the angular distance). The PSO model can reproduce many structural properties of real networks: clustering, small-worldness (concurrent low characteristic path length and high clustering), node degree heterogeneity with power-law degree distribution and rich-clubness. However, being the nodes uniformly distributed over the angular coordinate, the model lacks a non-trivial community structure.

The nonuniform PSO (nPSO) model[3,4] is a recently introduced generative model for realistic networks that is based on growing soft random geometric graphs with tailored community organization in the hyperbolic space. It is a generalization of the PSO model that exploits a nonuniform distribution of nodes over the angular coordinate in order to generate networks characterized by communities, with the possibility to tune their number, size and mixing property. In this study, we adopted a Gaussian mixture distribution of angular coordinates, with communities that emerge in correspondence of the different Gaussians, and the parameter setting suggested in the original studies[3,4]. Given the number of components $C$, they have means equidistantly arranged over the angular space, $\mu_i = \frac{2\pi}{C} \cdot (i-1)$, the same standard deviation fixed to 1/6 of the distance between two adjacent means, $\sigma_i = \frac{1}{6} \cdot \frac{2\pi}{C}$, and equal mixing proportions, $\rho_i = \frac{1}{C}$ $(i = 1 \dots C)$. The community memberships are assigned considering for each node the component whose mean is the closest in the angular space. The other parameters of the model are: *N*, the number of nodes; *m*, around half of the average node degree; *T*, the network temperature, inversely related to the clustering; *γ*, the exponent of the power-law degree distribution. Given the input parameters (*N*, *m*, *T*, *γ*, *C*), the nPSO model provides in output: the adjacency matrix of the network; the geometrical coordinates of the nodes in the hyperbolic disk, the community memberships of the nodes; the pairwise hyperbolic distances (geodesics) between the nodes. For details on the generative procedure please refer to the original studies[3,4]. The MATLAB code is publicly available at the GitHub repository: https://github.com/biomedical-cybernetics/nPSO_model.

## Structural connectomes data

The dataset includes tractography-based connectivity matrices of 614 healthy individuals (Male=230, Female=384) generated by the enhanced Nathan Kline Institute-Rockland Sample (NKI-RS; fcon_1000.projects.nitrc.org/indi/enhanced/)[15]. Streamline count adjacency matrices were constructed by counting the NOS that terminated in each region of interest of the Yeo network functional parcellation (114 cortical nodes)[16]. The whole dataset (n=614) was used to assess gender differences at the brain network level. In addition, from this dataset we also extracted a subset of n=438 connectivity matrices of individuals in two different age ranges: [7, 30] years old (n=223) and [55, 85] years old (n=215). Further details on this dataset are available in[13], which is the study that processed and provided us the connectomes.

The NOS weights represent connection strengths, however our computation requires weights to represent distances between the nodes. Longer white matter projections are more expensive in terms of their material and energy costs, thus making brain regions that are spatially close more likely to be connected[17]. Following this rationale, two brain regions connected by a higher number of streamlines tend to be at lower distance. Therefore, for every edge (i,j), the weight has been reversed according to the following formula:

$$w^*(i,j) = \frac{1}{1 + w(i,j)}$$

where $w(i,j)$ is the original weight (NOS) between the adjacent nodes i and j and $w^*(i,j)$ represent the reversed weight which we consider for our brain connectomic analysis.

## Hardware and software
MATLAB code has been used for all the simulations, carried out partly on a workstation under Windows 8.1 Pro with 512 GB of RAM and 2 Intel(R) Xenon(R) CPU E5-2687W v3 processors with 3.10 GHz, and partly on the ZIH-Cluster Taurus of the TU Dresden.

## Funding
Work in the CVC laboratory was supported by the independent research group leader starting grant of the Technische Universität Dresden.

## Author Contributions
CVC conceived the geometrical congruence and the content of the study. Both the authors contributed to design computational experiments, figures and items. AM implemented the code for the computation of GEO, GSP, pTSP and GRE; CVC prototyped the code for Figures 2 and 3; AM finalized the computational analysis and realized figures and items. Both the authors analyzed and interpreted the results. CVC wrote the main section of the article, and AM corrected it. AM wrote the Methods section and Supplementary Information Section, and CVC corrected it. CVC planned, directed and supervised the study.

## Competing interests
The authors declare no competing financial interests.

## Acknowledgements
We thank the BIOTEC System Administrators for their IT support, Gloria Marchesi and Claudia Matthes for the administrative assistance, the Centre for Information Services and High Performance Computing (ZIH) of the TU Dresden.

Figures

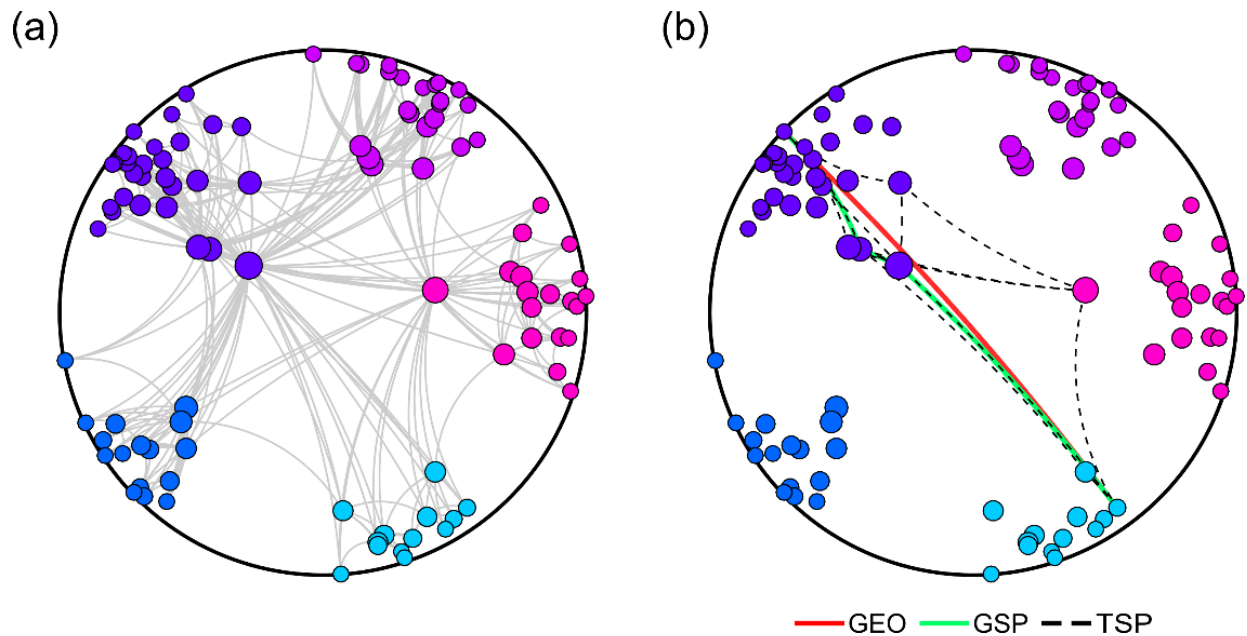

**Figure 1. Visualization of geodesics, geometrical and topological shortest paths in hyperbolic space.**
A nPSO network has been generated with parameters N = 100 (network size), m = 4 (around half of average degree), T = 0.1 (temperature, inversely related to the clustering coefficient), γ = 2.5 (power-law degree distribution exponent) and C = 5 (number of communities). **(a)** Representation of the nPSO network in the hyperbolic disk: the links are in grey colour and follow the hyperbolic geodesics, the nodes are coloured by community membership and their size is proportional to the logarithm of the degree. **(b)** The panel highlights in red colour the hyperbolic geodesic (GEO) between two specific nonadjacent nodes in the network, in black colour the links involved in all the possible topological shortest paths (TSP) between these two nodes and in green colour the geometrical shortest path (GSP).

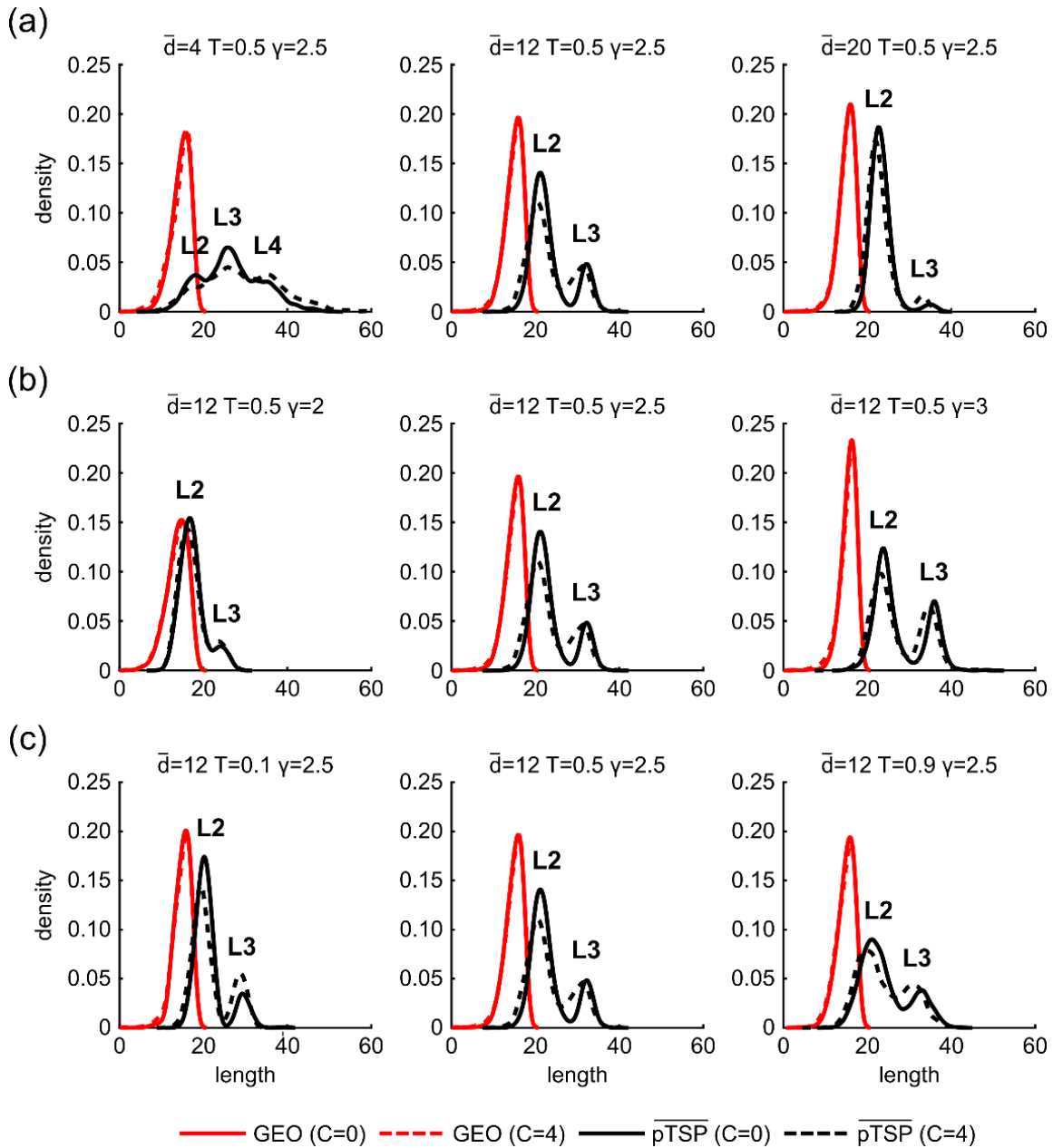

**Figure 2. Comparison of GEO and $\overline{pTSP}$ distributions in hyperbolic networks.**
For each subplot, the nPSO network has been generated with parameters N = 100, C = [0, 4] and values of $\bar{d}$, T, γ as indicated on top of each subplot. In particular: **(a)** we fixed T = 0.5, γ = 2.5 and varied $\bar{d}$ = [4, 12, 20]; **(b)** we fixed $\bar{d}$ = 12, T = 0.5 and varied γ = [2, 2.5, 3]; **(c)** we fixed $\bar{d}$ = 12, γ = 2.5 and varied T = [0.1, 0.5, 0.9]. The different results for C = 0 and C = 4 are shown within each subplot with a solid and dashed line respectively. For a given network, we have computed the geodesics (GEO) and the $\overline{pTSP}$ for all pairs of nonadjacent nodes. Then, for both GEO and $\overline{pTSP}$ we have estimated by kernel-density the probability density function, which is reported in the subplots (GEO in red, $\overline{pTSP}$ in black): x-axis represents the length of the path (GEO or $\overline{pTSP}$) and y-axis the density function. For $\overline{pTSP}$ we also highlight the peaks of the distribution that correspond to topological shortest paths of length 2, 3 and 4.

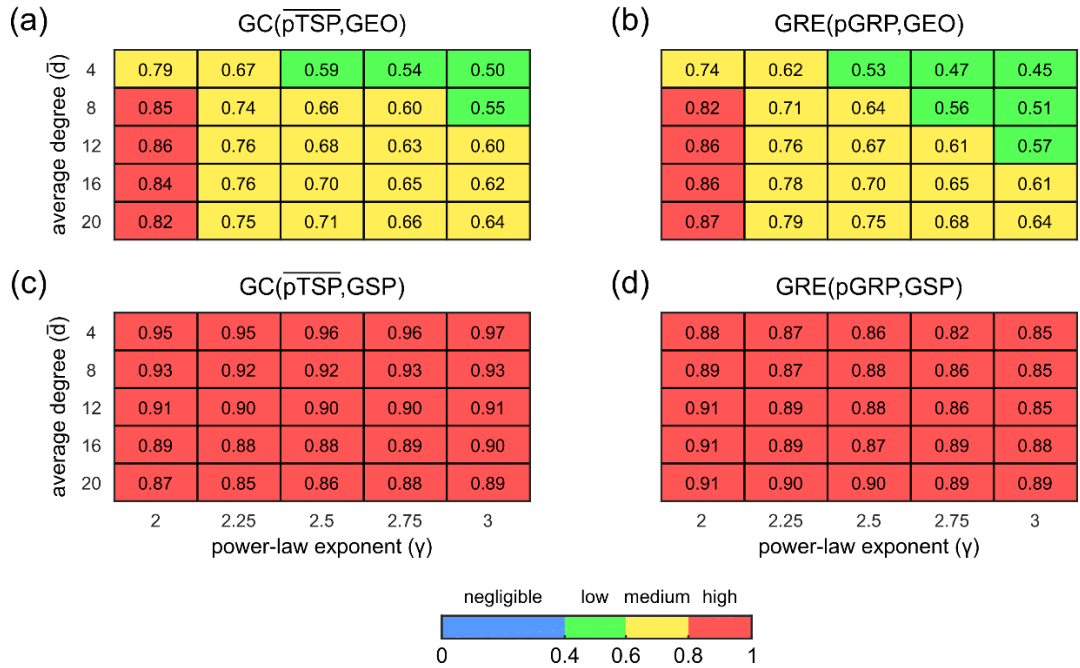

**Figure 3. GC and GRE evaluation on hyperbolic networks.**
nPSO networks have been generated with parameters N = 100, $\bar{d}$ = [4, 8, 12, 16, 20], T = 0.1, γ = [2, 2.25, 2.5, 2.75, 3] and C = 4. For each combination of parameters, 10 networks have been generated. For each network we have computed: **(a)** GC($\overline{pTSP}$, GEO), **(b)** GRE(pGRP, GEO), **(c)** GC($\overline{pTSP}$, GSP) and **(d)** GRE(pGRP, GSP). Each heatmap reports the mean value (over 10 network realizations) of the respective network measure for each combination of $\bar{d}$ and γ in the nPSO generative model.

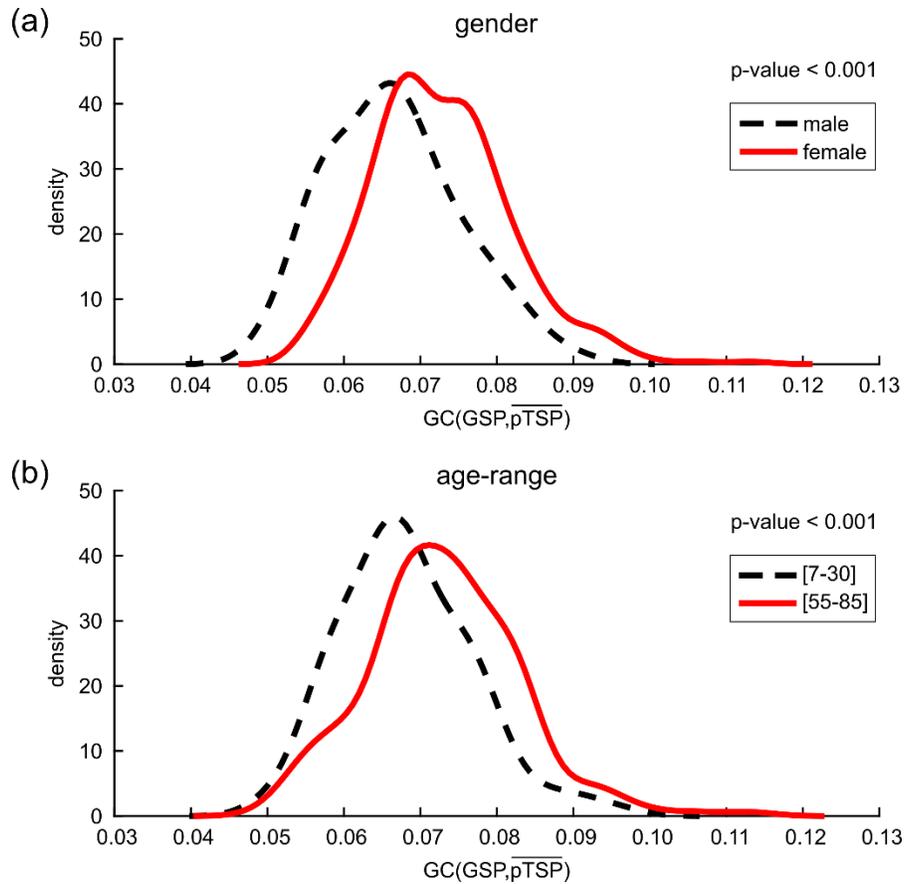

**Figure 4. GC as marker for gender and age discrimination in structural connectomes.**
We have analysed a dataset of structural connectomes with gender and age annotation (see Methods for details). For each network, we have computed the GC(GSP, $\overline{pTSP}$). **(a)** We divided the connectomes in two groups related to male and female subjects, for each group we have estimated by kernel-density the probability density function of GC, which is reported in the subplot (male in black dashed line, female in red solid line): x-axis represents the GC(GSP, $\overline{pTSP}$) and y-axis the density function. The p-value of the Mann-Whitney test shows that GC can significantly discriminate between male and female connectomes (p-value<0.001). **(b)** We have repeated the same analysis as in (a), but considering two groups of connectomes related to two age-ranges of the subjects: [7-30] (black dashed line) and [55-85] (red solid line). The p-value of the Mann-Whitney test shows that GC can significantly discriminate between connectomes of young and elderly subjects (p-value<0.001).

# Supplementary Information

**Suppl. Algorithm 1. Pseudocode to find all the paths of a given length between two nodes.**

---

**INPUT**
A = adjacency list
s = source node
t = target node
L = path length

**OUTPUT**
paths = list of paths, where each path is a list of L+1 nodes corresponding to the sequence of nodes involved in the path from s to t

---

**function** paths = find_paths(A, s, t, L)

paths = list()                          # initialize paths to empty list
path = list()                           # initialize path to empty list

paths = find_paths_rec(A, s, t, L, paths, path)        # start recursion from the source node

---

**function** paths = find_paths_rec(A, u, t, L, paths, path)

l = length(path)-1                      # current path length

if (u!=t) & (u not in path) & (l+1<L)   # if u is not the target and not in the path, and L is not reached
    path.append(u)                   # add node u to the path
    for v in A[u]                    # for each neighbour of the node u
        paths = find_paths_rec(A, v, t, L, paths, path)       # continue recursion
    end
elseif (u==t) & (l+1==L)                # if u is the target and L is reached
    path.append(t)                   # add target node to the path
    paths.append(path)               # store the path
end

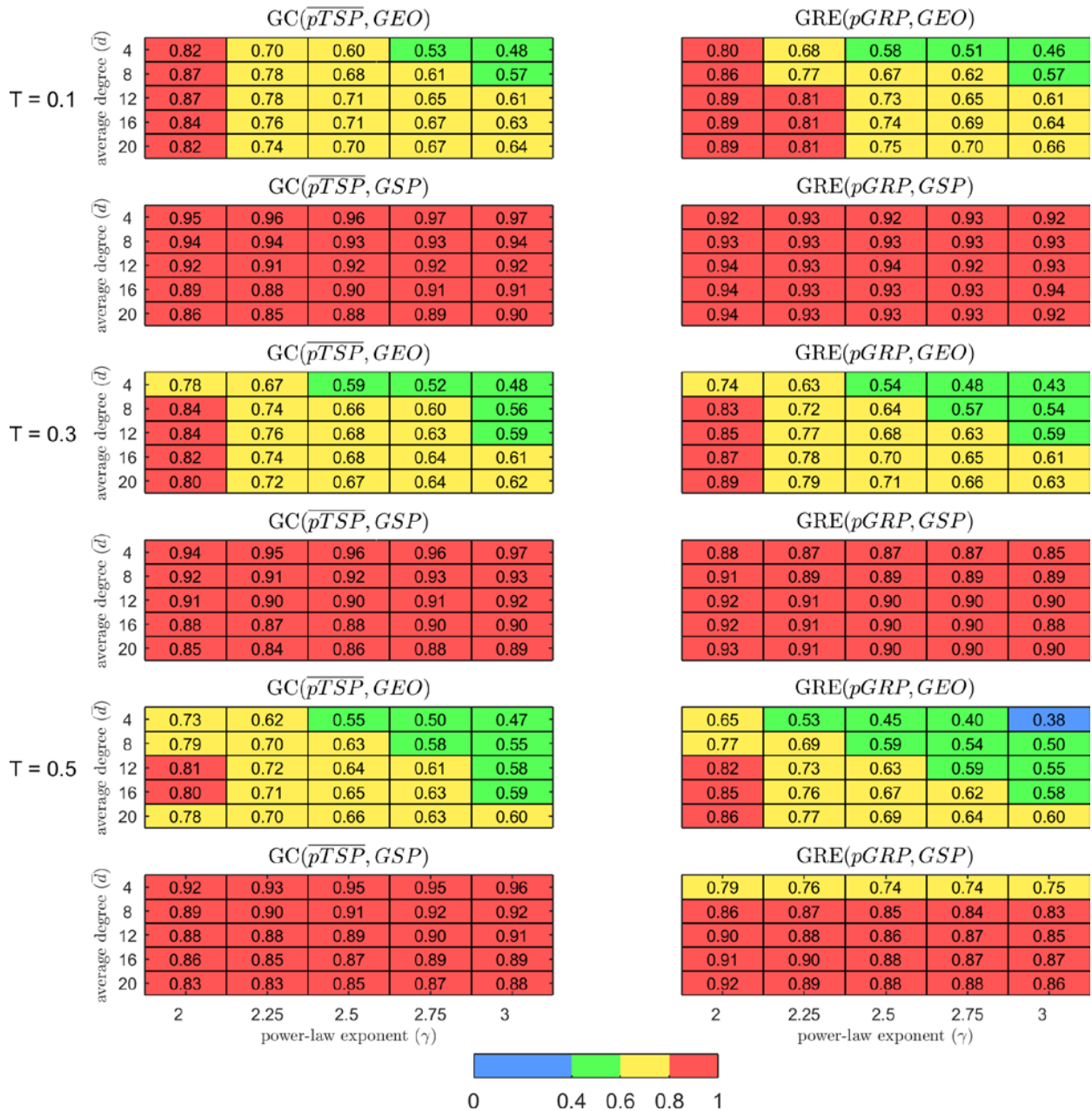

**Suppl. Figure 1. GC and GRE evaluation on nPSO networks (C = 0, N = 100).**
nPSO networks have been generated with parameters N = 100, $\bar{d}$ = [4, 8, 12, 16, 20], T = [0.1, 0.3, 0.5], γ = [2, 2.25, 2.5, 2.75, 3] and C = 0. For each combination of parameters, 10 networks have been generated. For each network we have computed: GC($\overline{pTSP}$, GEO), GRE(pGRP, GEO), GC($\overline{pTSP}$, GSP) and GRE(pGRP, GSP). For each value of T, indicated on the left, each heatmap reports the mean value (over 10 network realizations) of the respective network measure for each combination of $\bar{d}$ and γ in the nPSO generative model.

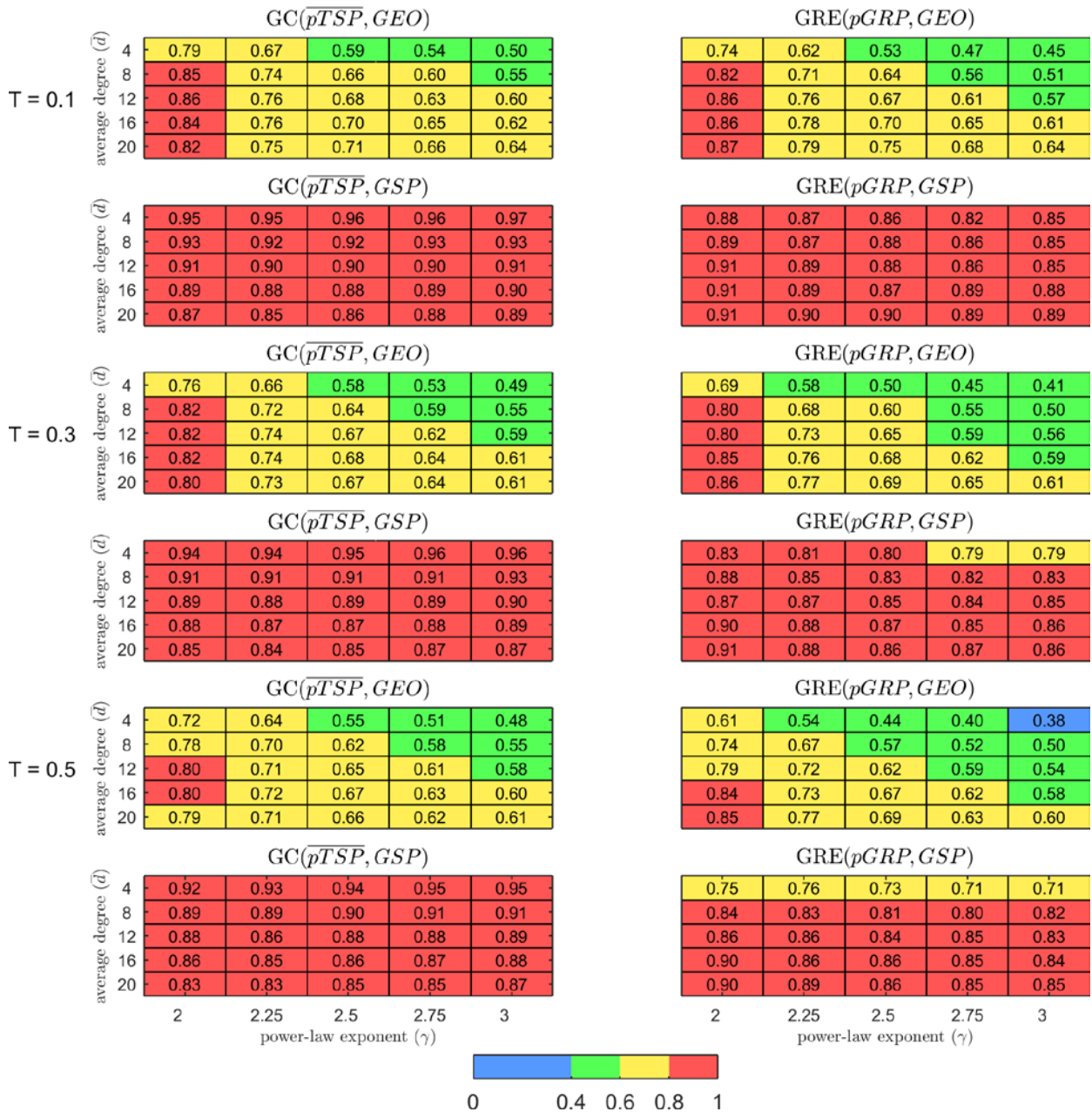

**Suppl. Figure 2. GC and GRE evaluation on nPSO networks (C = 4, N = 100).**
nPSO networks have been generated with parameters N = 100, $\bar{d}$ = [4, 8, 12, 16, 20], T = [0.1, 0.3, 0.5], γ = [2, 2.25, 2.5, 2.75, 3] and C = 4. For each combination of parameters, 10 networks have been generated. For each network we have computed: GC($\overline{pTSP}$, GEO), GRE(pGRP, GEO), GC($\overline{pTSP}$, GSP) and GRE(pGRP, GSP). For each value of T, indicated on the left, each heatmap reports the mean value (over 10 network realizations) of the respective network measure for each combination of $\bar{d}$ and γ in the nPSO generative model.

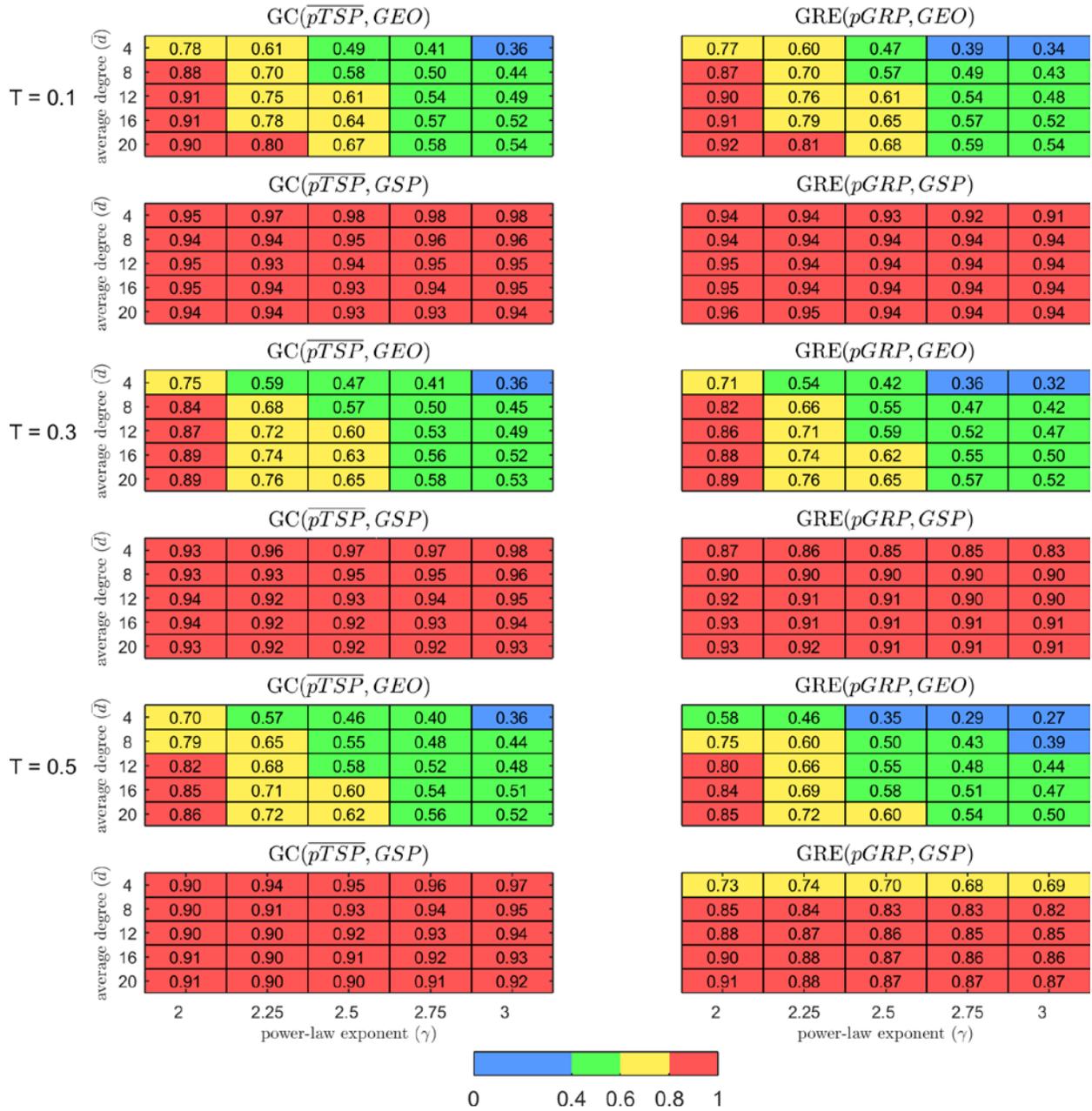

**Suppl. Figure 3. GC and GRE evaluation on nPSO networks (C = 0, N = 1000).**
nPSO networks have been generated with parameters N = 1000, $\bar{d}$ = [4, 8, 12, 16, 20], T = [0.1, 0.3, 0.5], γ = [2, 2.25, 2.5, 2.75, 3] and C = 0. For each combination of parameters, 10 networks have been generated. For each network we have computed: GC($\overline{pTSP}$, GEO), GRE(pGRP, GEO), GC($\overline{pTSP}$, GSP) and GRE(pGRP, GSP). For each value of T, indicated on the left, each heatmap reports the mean value (over 10 network realizations) of the respective network measure for each combination of $\bar{d}$ and γ in the nPSO generative model.

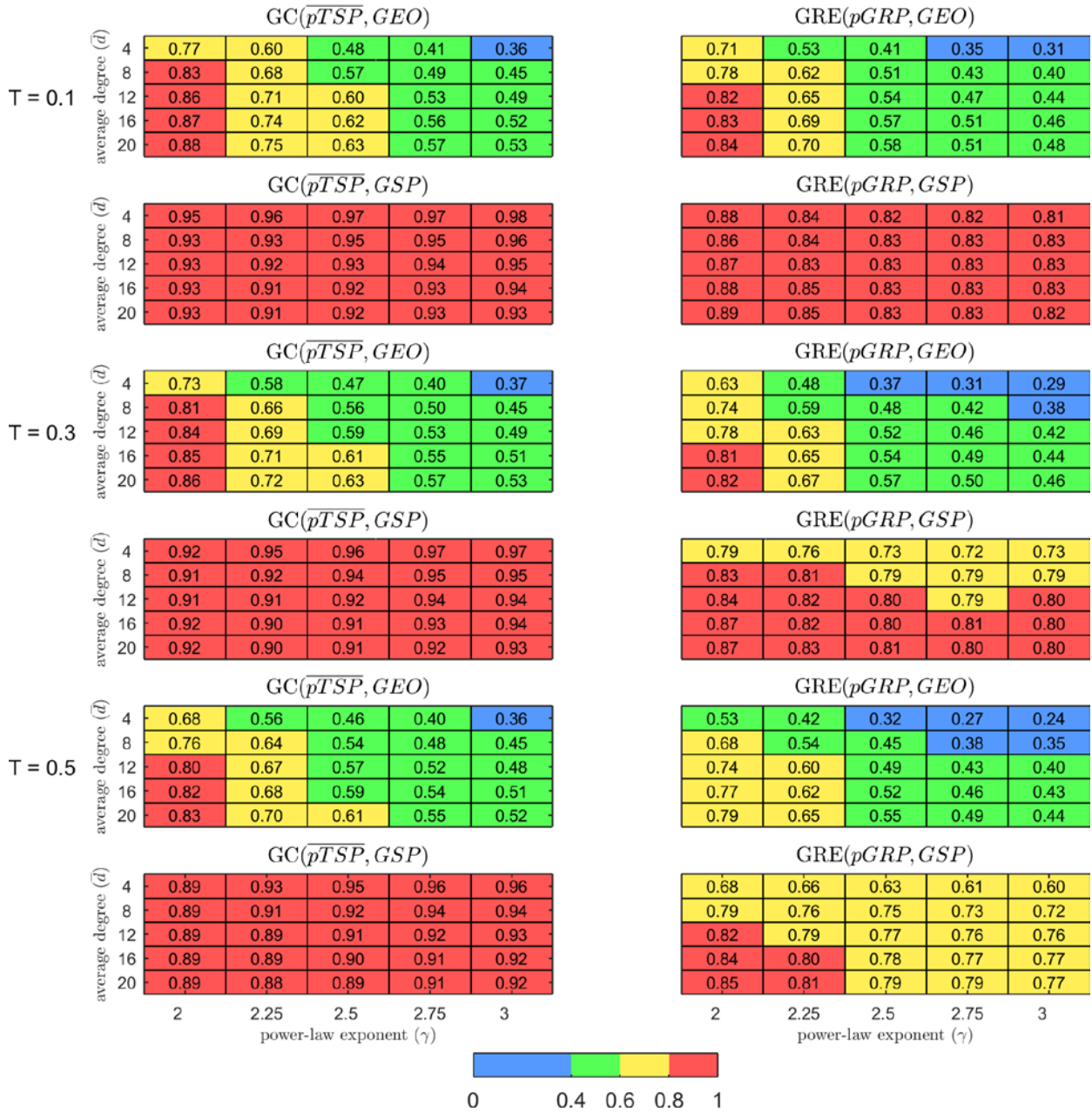

**Suppl. Figure 4. GC and GRE evaluation on nPSO networks (C = 4, N = 1000).**
nPSO networks have been generated with parameters N = 1000, $\bar{d}$ = [4, 8, 12, 16, 20], T = [0.1, 0.3, 0.5], γ = [2, 2.25, 2.5, 2.75, 3] and C = 4. For each combination of parameters, 10 networks have been generated. For each network we have computed: GC($\overline{pTSP}$, GEO), GRE(pGRP, GEO), GC($\overline{pTSP}$, GSP) and GRE(pGRP, GSP). For each value of T, indicated on the left, each heatmap reports the mean value (over 10 network realizations) of the respective network measure for each combination of $\bar{d}$ and γ in the nPSO generative model.